\documentclass[epjST]{svjour}
\usepackage{graphics}
\usepackage{epsfig}
\usepackage{graphicx}
\usepackage{amsmath}
\usepackage{amsfonts}

\begin{document}
\title{Energy of strongly attractive Bose-Fermi mixtures}
\subtitle{A comparison between T-matrix and Quantum Monte Carlo results}
\author{Andrea Guidini\inst{1}\fnmsep\thanks{\email{andrea.guidini@unicam.it}} \and Elisa Fratini\inst{2} \and Gianluca Bertaina\inst{3} \and Pierbiagio Pieri\inst{1} }
\institute{ School of Science and Technology, Physics Division, 
University of Camerino, Via Madonna delle Carceri 9, I-62032 Camerino, Italy \and The Abdus Salam International Centre for Theoretical Physics, 34151 Trieste, Italy  \and Dipartimento di Fisica, Universit\`a degli Studi di Milano, via Celoria 16, I-20133 Milano, Italy }
\abstract{
We discuss how approximate theories for Bose-Fermi mixtures recover in the molecular limit the expected expression for Fermi-Fermi mixtures of molecules and unpaired fermions. In particular, we compare the energy of the system resulting from a T-matrix diagrammatic approach with that obtained with Variational and Fixed-Node Diffusion Quantum Monte Carlo methods.
} %end of abstract
\maketitle
\section{Introduction}
\label{introduction}
Bose-Fermi mixtures with a tunable boson-fermion attraction have been object of active theoretical and experimental investigation over the last few years (see \cite{Yu11,Lud11,fra12,And12,ber13,fra13,Sog13,gui2014} and~\cite{Wu11,Wu12,Park12,Heo12,Cum13,Blo13}  for recent theoretical and experimental works, respectively). Previous theoretical studies of these systems have shown that for a sufficiently strong attraction between fermions and
bosons, the boson condensation is completely suppressed in mixtures where the boson density $n_{\rm B}$ does not exceed the fermion density $n_{\rm F}$.
This complete suppression of condensation occurs even at zero temperature, and is associated with pairing of bosons with fermions into composite
fermions.
 This kind of evolution has been studied already by us with a T-matrix diagrammatic formalism~\cite{fra10,fra12,fra13,gui2014} and with the Fixed-Node Diffusion Monte Carlo method~\cite{ber13}.

In the strongly attractive regime, for $n_{\rm F}>n_{\rm B}$, the system can be represented as a mixture of two different species of fermions: molecules and unpaired fermions. 
In \cite{gui2014} we have shown how the molecular limit emerges from a T-matrix approach for Bose-Fermi mixtures. There, we focused on the momentum distribution function of the two species. Here, we present results for the energy and chemical potentials that were not reported before, and compare them with Variational (VMC) and Fixed-Node Diffusion (FN-DMC) Quantum Monte Carlo calculations, as well as with the beyond-mean-field equation of state for a repulsive Fermi-Fermi mixture obtained recently in Ref.~\cite{fra2014} within second-order perturbation theory.
\section{Formalism}
\label{formalism}
We consider a homogeneous mixture, composed by single-component fermions and bosons of equal masses $m_{\rm F}=m_{\rm B}$ with densities $n_{\rm F}$ and $n_{\rm B}$, respectively, at temperature $T$=0. The boson-fermion interaction is assumed to be tuned by a broad Fano-Feshbach resonance. Under this condition, the boson-fermion interaction can be adequately described by an attractive point-contact potential, with the interaction parametrized in terms of the boson-fermion scattering length $a$.
The repulsion between bosons, which stabilizes the system in the resonance region, is not necessary in the strongly attractive (molecular) limit of interest
to the present paper, and is not included in the T-matrix calculations. In the Quantum Monte Carlo (QMC) calculations, on the other hand, for consistency with Ref.~\cite{ber13} we set $\zeta \equiv k_{\rm F} a_{\rm BB} = 1$, where $a_{\rm BB}$ is the boson-boson (positive) scattering length and $k_{\rm F}$ is the Fermi momentum associated to the fermionic density $n_{\rm F}=k_{\rm F}^3/(6\pi^2)$. We use the dimensionless coupling parameter $g=(k_{\rm F} a)^{-1}$ to describe the strength of the interaction and consider the strongly attractive regime, where $a$ is small and positive and $g \gg 1$. In this limit the system can be described in terms of excess unpaired fermions and composite fermions, i.e. boson-fermion pairs with a binding energy $\epsilon_0=1/(2 m_r a^2)$, $m_r$ being the reduced mass $m_r=\frac{m_{\rm B}m_{\rm F}}{m_{\rm B}+m_{\rm F}}$ where $m_{\rm B}$ and $m_{\rm F}$ are the boson and fermion masses, respectively (and we set $\hbar$=1 throughout).
\subsection{T-matrix equations}
In general the bosonic and fermionic chemical potentials $\mu_{\rm B}$ and $\mu_{\rm F}$ need to be determined by solving numerically the T-matrix equations (see~\cite{fra12}) for given values of the densities and of the boson-fermion scattering length. Analytic expressions can be derived in the molecular limit by expanding the T-matrix equations in inverse powers of the bosonic chemical potential \cite{gui2014}. Actually, for $g \gg 1$ and $n_{\rm B} \leq n_{\rm F}$, $\mu_{\rm B}$ approaches minus the binding energy, that is the largest energy scale of the system in the molecular limit ($\epsilon_0 \gg E_{\rm F}$, $\mu_{\rm B} \sim - \epsilon_0$, where $E_{\rm F}=k_{\rm F}^2/(2m_{\rm F})$ is the Fermi energy associated to the fermionic density). The asymptotic equations for the chemical potentials are then given by: 
\begin{eqnarray}
\label{mub_sc}
\mu_{\rm B} &=& \frac{(6 \pi ^2 n_{\rm B})^{2/3}}{2 M}+\frac{4\pi a}{m_r}(n_{\rm F}-2n_{\rm B})-\frac{[6 \pi ^2(n_{\rm F}-n_{\rm B})]^{2/3}}{2m_{\rm F}}-\epsilon_0, \\
\label{muf_sc}
\mu_{\rm F} &=& \frac{[6\pi^2(n_{\rm F}-n_{\rm B})]^{2/3}}{2m_{\rm F}}+\frac{4 \pi a}{m_r}n_{\rm B}.
\end{eqnarray}
Note that in the above equations~the fermion-molecule repulsion (described by the last term of Eq.~(2), and contained also in Eq.~(1)) is treated only approximately, with a fermion-molecule scattering length 
$a_{\rm FM} = (1 + m_{\rm F}/m_{\rm B})^2/(1/2 + m_{\rm F}/m_{\rm B}) a$, as it can be seen by introducing the reduced mass for a molecule and one fermion. This value corresponds to a kind of Born approximation to the exact values calculated in~\cite{skorniakov}, for $m_{\rm B}=m_{\rm F}$, and in \cite{iskin}, in the general case.

The energy is calculated by performing an integration of the bosonic chemical potential over the bosonic density $n_{\rm B}$ from 0 up to a chosen value $\bar{n}_{\rm B}$ keeping constant the fermionic density $\bar{n}_{\rm F}$ as it follows:
\begin{equation}
\label{energy_eq}
E=N_{\rm F} E_{\rm FG} + V \int _0 ^{\bar{n}_{\rm B}} {\rm d}n_{\rm B} \phantom{q} \mu_{\rm B}(n_{\rm B},\bar{n}_{\rm F}),
\end{equation}
where $V$ is the volume occupied by the system, $E_{\rm FG}=\frac{3}{5}E_{\rm F}$ is the energy per particle of the non-interacting Fermi gas, $N_{\rm F}$ is the number of fermions and the bosonic chemical potential $\mu_{\rm B}$ is determined by solving numerically the T-matrix equations~\cite{fra12}. In principle, the energy could be obtained also from the (bosonic or fermionic) single-particle Green's function (cf., e.g.  Eq.~(7.27) of \cite{fetter}). The ensuing integrals over momentum and frequency are however slowly convergent and hard to be tackled numerically. For this reason, we preferred to calculate the energy through a numerical integration of the data for the bosonic chemical potential. 
\subsection{Quantum Monte Carlo}
In the strongly attractive regime the trial wave function used to perform both Variational (VMC) and Fixed-Node Diffusion Quantum Monte Carlo (FN-DMC) is the product of a symmetric Jastrow function and a Slater determinant for the molecules and the unpaired fermions, which satisfies the fermionic antisymmetry condition. The details of the simulations are the same as in \cite{ber13}. We observe that the trial wave function is not required to be symmetric under the exchange of bosons, when used just for the calculation of the energy. On the contrary, in \cite{gui2014} a suitable symmetrization of the molecular trial wave function has been introduced to calculate the momentum distributions.
\subsection{Mean-Field and Second-Order Perturbation Theory}
At the mean-field level the equation of state of the Fermi-Fermi mixture of repulsive fermions resulting in the molecular limit of a Bose-Fermi mixture is:
\begin{equation}\label{meanfield}
E^{\rm mf}=N_{\rm F}E_{\rm FG}\left[(1-x)^{\frac{5}{3}}+\frac{1}{\tilde M}x^{\frac{5}{3}}+\frac{10}{9\pi}\frac{1+\tilde M}{\tilde M} x(1-x) \frac{\alpha}{g}\right],
\end{equation}
where $\tilde{M}=M/m_{\rm F}$ with $M=m_{\rm B}+ m_{\rm F}$, and $x=N_{\rm B}/N_{\rm F}$ is the concentration, which can be equivalently written in terms of the number of molecules $N_{\rm M}$ and unpaired fermions $N_{\rm UF}$ as $x=N_{\rm M}/(N_{\rm M}+N_{\rm UF})$.
The last term of Eq.~\ref{meanfield} describes the repulsion between composite and unpaired fermions, where $\alpha$ is the ratio between the fermion-molecule and boson-fermion scattering lengths. The solution of the scattering problem between a molecular and an unpaired fermion yields the value $\alpha=1.18$ \cite{skorniakov}, while if one integrates in Eq.~(\ref{energy_eq}) the strong-coupling expression of $\mu_{\rm B}$ given by Eq.~(\ref{mub_sc}) one gets Eq.~(\ref{meanfield}) (minus the binding energy) with the Born approximation value $\alpha=8/3$.

The equation of state for mass-imbalanced repulsive fermionic mixtures expanded up to the second order in the interaction parameter $g$ has been calculated in \cite{fra2014}, and is given by:
\begin{equation}\label{secondord}
E^{(2)}=E^{{\rm mf}}+N_{\rm F}E_{\rm FG}\frac{1}{2^{10/3}\pi^2}\left(\frac{1+\tilde M}{\tilde M}\right)^2 I(x,\tilde M) \left(\frac{\alpha}{g}\right)^2,
\end{equation}
where $I(x,\tilde M)=1.31396$ for a Bose-Fermi mixture with $m_{\rm F}=m_{\rm B}$ and $x=0.175$.
\begin{figure}[!htbp]
\begin{center}
\includegraphics[scale=0.6]{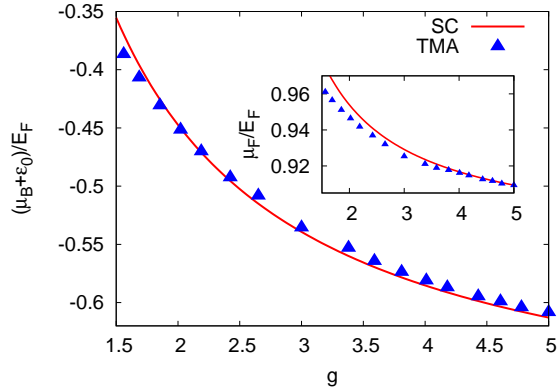} \caption{Bosonic chemical potential $\mu_{\rm B}$ (minus the binding contribution $-\epsilon_0$) as a function of the Bose-Fermi attraction $g$ for a mixture with $m_{\rm B}=m_{\rm F}$ and $x=0.175$. Inset: fermionic chemical potential $\mu_{\rm F}$ as a function of the Bose-Fermi attraction $g$. Triangles: T-matrix results; full lines: asymptotic expressions (\ref{mub_sc}-\ref{muf_sc}).}
\label{pot_chim_sc_fig}
\end{center}
\end{figure}
\section{Results}
\label{results}
In Figure \ref{pot_chim_sc_fig} we present a comparison between the chemical potentials obtained from the full numerical solution of the T-matrix equations (blue triangular points) and those obtained using the strong-coupling expressions of Eqs.~(\ref{mub_sc}-\ref{muf_sc}) (solid red curves) as functions of the Bose-Fermi attraction $g$. 
We have considered a mass balanced Bose-Fermi mixture with a concentration $x=n_{\rm B}/n_{\rm F}=0.175$.
Increasing the interaction strength, both the bosonic and fermionic numerical chemical potentials approach the corresponding asymptotic expressions derived in the molecular limit for $g \gg 1$.

\begin{figure}[!bp]
\begin{center}
\includegraphics[scale=0.6]{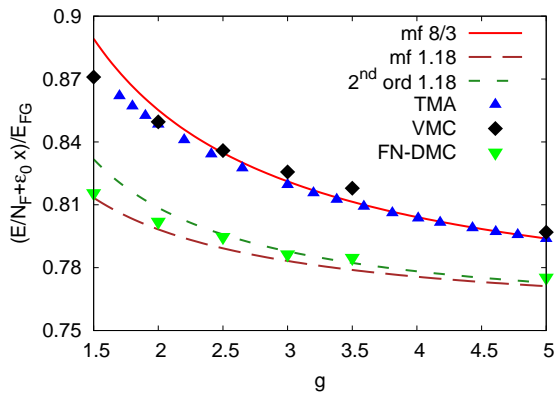}\caption{Energy of a Bose-Fermi mixture with $m_{\rm B}=m_{\rm F}$ and $x=0.175$ vs.~$g$ (with the contribution of the bare binding energy of the molecules subtracted). Full line: mean-field energy (\ref{meanfield}) with $\alpha$=8/3. Long-dashed line:  mean-field energy (\ref{meanfield}) with $\alpha$=1.18. Short-dashed line: Eq.~(\ref{secondord}) with $\alpha$=1.18.}
\label{energy_fig}
\end{center}
\end{figure}

The corresponding results for the energy are shown in Figure \ref{energy_fig}.   
We observe a good agreement between the VMC and T-matrix results (TMA), both approaching the mean-field equation of state (\ref {meanfield}) with a fermion-molecule scattering length $a_{\rm FM}=8/3a$. 
This shows that while the T-matrix approach and VMC with trial wave function given by Eq.~(2) of Ref.~\cite{ber13} capture the internal structure of the molecule forming in the medium, they are less accurate in describing the interaction between the molecules and the unpaired fermions, both yielding the Born approximation value 8/3$a$ for the fermion-molecule scattering length instead of the exact value 1.18$a$~\cite{skorniakov}. 

The FN-DMC method performs an imaginary-time evolution of the initial trial wave function (Eq.~(2) of Ref.~\cite{ber13}), thus projecting onto the ground-state with the same nodal surface (Fixed-Node approximation). This effectively optimizes the many-body correlations and improves on the VMC results. Actually the FN-DMC energy approaches the perturbative curves (mean-field and second order) with the exact fermion-molecule scattering length 1.18$a$ in the strong-coupling limit. In particular, for large values of $g$ the FN-DMC results seem to approach the equation of state obtained with second-order perturbation theory.

\section{Concluding remarks}
\label{conclusion}
In summary we have shown how, within a T-matrix diagrammatic approach, a Fermi-Fermi mixture emerges effectively from a Bose-Fermi mixture for sufficiently strong
attraction.
In this limit, we have derived simple expressions for the chemical potentials that compare well with the full T-matrix results. Furthermore, we have calculated the energy of such a mixture for $m_{\rm B}=m_{\rm F}$ and shown that the T-matrix energy is in good agreement with the energy calculated within a VMC framework, both of them approaching the mean-field value of the energy of a repulsive Fermi-Fermi mixture with fermion-molecule scattering length $a_{\rm FM}=8/3a$. The correct fermion-molecule scattering length $a_{\rm FM}=1.18a$ is instead recovered by  FN-DMC. In this case, the inclusion of the second-order perturbative correction to the energy of a repulsive Fermi-Fermi mixture recently found in \cite{fra2014} seems to slightly improve the agreement with the asymptotic expression for large $g$.  Our calculations focused on the case of equal masses to limit the computational effort of the QMC simulations. The main conclusions of the present work should however be more general.  
We believe in particular that, also for different masses, the VMC and T-matrix results should agree  in the molecular limit and recover the Born approximation for the fermion-molecule scattering length, the correct result being obtained only by FN-DMC.  Only future QMC simulations for different mass ratios will however be able to confirm or disprove our physical expectations in a definite way.

\end{document}